\documentclass[11pt]{article}

\usepackage{amssymb,amsthm,amscd, amsbsy, array}
 \usepackage{amsmath,bm}
\def\vq{{\bf q}}
\def\vr{{\br}}
\def\lf{\left(}
\def\rg{\right)}
\def\lq{\left[}
\def\rq{\right]}
\def\lgr{\left\{}
\def\rgr{\right\}}
\newcommand{\half}{{\scriptstyle{\frac{1}{2}}}}

\def\cA{{\cal A}}
\def\cE{{\cal E }}
\def\cR{{\cal R}}

\def\cL{{\cal L}}

\def\vB{{\bB}}
\def\vE{{\vec{E}}}
\def\bB{{\vec{B}}}
\def\br{{\vec{r}}}
\def\bE{{\vE}}
\def\bk{{\vec{k}}}

\def\vTheta{{\vec{\Theta}}}

\def\p{{\partial}}
\def\vx{{\vec x}}

\def\vp{{\vec p}}

\def\vk{{\vec k}}

\def\beq{\begin{equation}}
\def\eeq{\end{equation}}
\def\beqa{\begin{eqnarray}}
\def\eeqa{\end{eqnarray}}
\def\nn{\nonumber}

\def\smallover#1/#2{\hbox{$\textstyle\frac{#1}{#2}$}} %
\newcommand{\const}{\mathop{\rm const}\nolimits}

\let\ssection=\section
\renewcommand{\section}{\setcounter{equation}{0}\ssection}
\def\vq{{\vec q}}
\def\vp{{\vec p}}

\def\vq{{\vec q}}
\def\vr{{\vec r}}

\def\vv{{\vec v}}

\def\vx{{\vec x}}

\def\vB{{\vec B}}

\def\vg{{\vec g}}

\begin{document}

\title{ Non-commutative mechanics and Exotic Galilean symmetry}

\author{
L. Martina\footnote{e-mail: Luigi.Martina@le.infn.it}
\\
Dipartimento di Fisica - Universit\`a del Salento
\\
and\\
Sezione INFN di Lecce. Via Arnesano, CP. 193\\
I-73 100 LECCE (Italy) 
}

\maketitle

\begin{abstract}
In order to derive a large set of Hamiltonian dynamical systems,
but with only first order Lagrangian, we resort to the formulation
in terms of Lagrange-Souriau 2-form formalism. A wide class of
systems derived in different phenomenological contexts are
covered.   The non-commutativity of the particle position
coordinates are a natural consequence. Some explicit examples are
considered.
\end{abstract}
\vspace{5mm}
\noindent


\newpage

\section{Introduction}

Recently interest in dynamical systems with non commuting
(not necessarily non canonical) variables stems from  Condensed
Matter Physics \cite{Xiao:2009rm}, Optics \cite{Bliokh:2009ek}
and  String Theory \cite{Szabo}. The applications include a Bloch
electrons \cite{Niu},  the Anomalous Hall Effect \cite{AHE}, the
Spin Hall Effect \cite{SpinHall}, and the Optical Hall effect
\cite{OptiHall}. For example,
\begin{itemize}

\item[a)]
The semiclassical equations of motion in the $\epsilon_n(\bk)$
energy band
 of a Bloch electron in a crystal solid read
\begin{eqnarray}
\dot{\br}=\displaystyle\frac{\p\epsilon_n(\bk)}{\p\bk}-\dot{\bk}\times\vTheta(\bk),\label{velrel}
\quad \dot{\bk}=-e\bE-e\dot{\br}\times\bB(\br). \label{Lorentz}
\end{eqnarray} The purely
momentum-dependent
$\Theta_i(\bk)=\epsilon_{ijl}\p_{\bk_j}\cA_l(\bk)$ is the
curvature associated to   the so-called  Berry connection $\cA$.
\item[b)]
The semiclassical equations that describe the spin-Hall effect
\label{spinHall} into semiconductors near the degenerate point
$\vk = 0$ of the valence band read  \beq \dot{\br}=\frac{\p
E_s(\bk)}{\p\bk}+\dot{\bk}\times\vTheta_s, \qquad \dot{\bk}=-e\vE,
 \eeq
 where $s=\pm \frac{1}{2},\pm  \frac{3}{2}$ is the spin helicity of the
 holes,  the  energy  is $E_s(\bk)= \frac{\hbar^2}{2 m}\lf A -B s^2 \rg k^2$
 and the Berry curvature due to the lattice structure is $\vTheta_s
 = s \lf 2 s^2 -\frac{7}{2}\rg \frac{\vk}{k^3}$.  The trajectories
 followed by opposite helicity holes separate during the motion,
 providing a tool for \textit{spintronic}.
\item[c)]
The \textit{optical Magnus}  and the \textit{optical Hall} effects
\cite{Bliokh:2009ek,OptiHall,SpinOptics} are described by the
approximate equations
\begin{eqnarray}
\dot{\vr}\approx{\vp}-\frac{s}{\omega}\,{\rm grad
}(\frac{1}{n})\times{\vp}, \qquad \dot{\vp}\approx
-n^3\omega^2{\rm grad }(\frac{1}{n}), \label{LinDuv}
\end{eqnarray}
where $s$ denotes the photon's spin, parametrizing a term which
 deviates the light's trajectory from the  predictions of ordinary
geometrical optics.
\item[d)] The motion of a Bogoliubov quasiparticle in a superfluid vortex   \cite{Bogo}
is described by
\begin{eqnarray}
M\dot{\vq}+ {\vec F} \times {\vr}=-\frac{\p h}{\p\vr}, \qquad
M\dot{\vr}=\frac{\p h}{\p\vq}, \label{Bogoeqmot}
\end{eqnarray}
where  the \textit{effective mass} $M_{ij}=\delta_{ij}-\frac{\p
A_i}{\p q^j}$ and \textit{magnetic field}  $F_{ij}=\frac{\p
A_i}{\p r^j}- \frac{\p A_j}{\p r^i}$ are defined in terms of a
gauge potential $\vec A = {\vec A}\lf \vq, \vr\rg$, and $h = H\lf
\vq, \vr \rg  $ is some Hamiltonian.
\item[e)]  The string theory inspired  dynamical system \cite{RoVe}
        \beq \dot{x}_i = \frac{p_i}{m} + \Theta_{i j} \frac{\p V}{\p x_j},
        \qquad  \dot{p}_i = - m \frac{\p V}{\p x_j}+ m \Theta_{i j} \frac{\p^2 V}{\p x_i \p
        x_j},\eeq
for the Kepler potential $V\propto r^{-1}$ in  a weak
non-commutative background  $\Theta$ predicts  a perihelion point
precession in the planetary motions, providing us with a measurable quantity to test
cosmological  models.
\end{itemize}
In all these models,
 the \textit{momentum} satisfies a first order ODE in terms of a
 (possibly  of the Lorentz-type) force.
 But the same structure appears in the equations for the position, which are
distinguished by an  unusual \textit{velocity relation}, as a
result  from the Berry phase contributions, associated with the
environment in which the particle moves,  or by a postulated
fundamental \textit{area scale}. Thus, the momentum and the
velocity may not be proportional.

 In the absence of any magnetic
field, for example, equations (\ref{spinHall}) reduce to those
used to explain the AHE observed in ferric materials \cite{AHE}:
the effect comes, entirely, from the \textit{anomalous velocity
term}. Let us emphasize that such a behavior has been advocated a
long time ago \cite{Dixon}. It has been shown indeed that the
Lorentz-equation follows from general principles, but the
momentum-velocity relation is model-dependent. Relations like
$\vp=m\dot{\vr}$ are indeed mere Ans\"atze, and it is not required
by any first principle.

 More recently,
similar features has been found in 2D systems for the so called
\textit{exotic} particles, i.e.,  ones which are associated with
the kinematical  two-parameter central extension of the planar
Galilei group \cite{LSZ,DH}. These models are, once again,
instrumental in describing the Fractional Quantum Hall Effect.
Those systems support a sort of \textit{duality } between the
magnetic field, $\bB(\br)$, and its analog in momentum space,
$\vTheta(\vp)$, called sometimes a dual magnetic field. By
consequence  the \textit{Poisson bracket } of the
\textit{position} variables they may no longer vanish, in analogy
with the components of the momentum in the presence of a magnetic
field.   In the relativistic case, similar, but more elaborated
constructions have been proposed \cite{Deriglazov,HP1}, which
however we do not discuss here.

 Even the question of consistency of the above effective models with
 the general principles of mechanics  is
legitimate, since they are mainly  derived  by some semiclassical
de-quantization procedure, which does not necessarily fit into the
framework of classical mechanics.  Thus, our primary goal below is
to prove that no new mechanics has to be invented~: all these
models fit perfectly into Souriau's presymplectic framework of
Classical Mechanics \cite{SSD}. The second aim is to formulate the
hamiltonian theory in that context, exploiting the
 supplementary
structure  encoded into the second central extension of the
Galilei group.

\section{ The Lagrange-Souriau 2-form }

 The  modern geometrical formulation \cite{SSD}    of the
calculus of variations, originated by Lagrange  and continued by
Cartan \cite{Cartan}, consists  in   mapping the Lagrangian
function $L=L(\vx,\vv,t)$, defined on the \textit{evolution space}
$TQ\times \mathbb{R} \rightarrow \mathbb{R}$, into the so-called
Cartan 1-form $\lambda$,  defined by   the crucial property \beqa
\int L(\gamma(t),\dot{\gamma},t)dt =
\int_{\widetilde{\gamma}}\lambda  \quad \textrm{with} \quad
\lambda=\frac{\p L}{\p p_i}dx^i+ \left(L-\frac{\p L}{\p
p_i}p_i\right)dt.\eeqa where
$\widetilde{\gamma}=(\gamma(t),\dot{\gamma},t)$ is the lifted
world-line in the tangent bundle
 of the evolution space $TQ\times \mathbb{R}$.
The exterior derivative  of the Cartan form provides us with a
closed \textit{Lagrange-Souriau} (\textit{LS}) 2-form $\sigma =
d\lambda$, which in general cannot be separated canonically into a
\textit{symplectic} and a \textit{Hamiltonian part} \cite{SSD}.
However, the associated Euler-Lagrange equations are still
expressed by determining the kernel of $\sigma$, i.e. by the
equation $ \sigma(\dot{\widetilde{\gamma}}, \cdot)=0 . $ If the
kernel is one-dimensional, the variational problem is called
regular. Otherwise  the singular case requires to resort to the
symplectic
 reduction techniques \cite{SSD,Cartan,FaddJ,HPLandau}. Thus  more general procedures
 have to be adopted to build such a
system and clarify their Hamiltonian structure.

Conversely, again following Souriau \cite{SSD}, a generalized
mechanical system is given by a  2-form $\sigma$,  which is closed
$d\sigma=0$ and  with constant rank $d$. Then, its kernel defines
an integrable foliation with $d$-dimensional leaves,  which can be
viewed as generalized solutions of the variational problem.
Moreover, by the Poincar\'e lemma,  $d\sigma=0$ implies  the local
existence of a Cartan $1$-form  $\lambda$. Rewriting it as
$\lambda=a_\alpha d\xi^\alpha$, one can plainly define  the
first-order Lagrangian function on the evolution space as \beq
\cL= a_\alpha\dot{\xi}^{\alpha} \quad \; \textrm{such that}\quad
\; \int \cL dt=\int\lambda .\label{1-order-action}\eeq  Thus, the
 above-mentioned models  do not have a usual Lagrange function
defined on the tangent bundle. Put in another way, the position
does not satisfy a second-order Newton equation.  The general
question  of the existence of a locally defined Lagrangian has
been discussed \cite{HPAJMP,Grigore}.

 Moreover,  if   the \textit{LS}
2-form $\sigma$
   can be split as
\beqa \sigma=\omega-dH\wedge dt, \label{sigma-split}\eeqa where
$\omega$ is a closed and regular 2-form on the \textit{phase
space} $TQ$ and $H$  is a Hamiltonian function on $TQ \times
\mathbb{R}$, than the equations of motion read \beqa
\omega\big(\dot{\widetilde{\gamma}}\big) =dH \label{sELeq}. \eeqa
Assuming that $\omega$ is  regular,  the inverse
$(\omega^{\alpha\beta})$ of the symplectic matrix
$\omega=\omega_{\alpha\beta}$ ( i.e.
$\omega^{\alpha\beta}\omega_{\beta\gamma}=
\delta^{\alpha}_{\gamma}$) yields the Hamilton equations
$\dot{\xi}_i=\{\xi_i, H\} $,
 through the
Poisson brackets  defined  by \beqa
\{f,g\}=\omega^{\alpha\beta}\p_\alpha f\p_\beta g.
\label{Poisson}\eeqa In the case of singularity, as said above,
symplectic reductions are needed.

Thus, in Souriau's framework, one can state the inverse problem of
the calculus of variations for given equations for a one-particle
system  in presence of a position-dependent force field $\vE$
only,  rewriting them as the set of 1-forms on $  TQ \times
\mathbb{R}$
\begin{eqnarray}
\alpha_1 = d \vr-\vp\; d t \qquad \alpha_2 = m \;d\vp-\vE \;dt ,
\end{eqnarray}
and view both of them as hyperplanes in the evolution space,
defined by the kernel of the one-forms $\alpha_i$. Notice that we
have introduced the kinetic momentum $\vp = m \vv$, and we will
refer to it in the sequel. Then, the simultaneous solutions
correspond to the intersection of these hyperplanes, described by
the kernel $ \sigma(\delta y,\cdot)=0 \;$ of
 exterior product $\;\sigma =\alpha_1 \wedge \alpha_2   $,  where $\wedge$ warrants the antisymmetry. In
presence of an electromagnetic field $\vE \lf \vr, \vp,
t\rg,\vB\lf \vr, \vp, t\rg$ acting on a charge $e$,  Souriau
\cite{SSD} generalized the previous simplest 2-form to
\begin{eqnarray}
\sigma &=& \lf m d \vv- e \vE dt \rg \wedge \lf d \vr- \vv dt \rg
+ e \vB \cdot d\vr \times d \vr, \label{Sem2}
\end{eqnarray}
where we have defined $ \lf d\vr \times d \vr \rg_{k} =
\frac{1}{2} \epsilon_{k i j} dx_i \wedge dx_j $. Then, the usual
equations of motion of a charged particle in the electromagnetic
field are seen to arise as the kernel of $\sigma$, together with
the regularity condition $d\sigma=0$   leading to the  the
homogeneous Maxwell equations.
These formulas can be readily generalized to the multi-particles
case.

Now, in the same spirit
we would like to write down a  Lagrangian 2-form for a particle of
mass $m$,
which is  subjected both to the electromagnetic field $\vE\lf
\vr,t \rg$ and $\vB\lf \vr,t \rg $,
and  to a peculiar environment, the local characteristics of which
may depend on the momentum.
\section{The (2+1)-dim Duval - Horv\'athy model }
The simplest example of such a situation is the \textit{exotic}
mechanical model in 2 space dimensions proposed in \cite{DH},
defined by a form generalizing (\ref{Sem2}) which can be split as
in (\ref{sigma-split}): precisely we introduce
\begin{equation}
\sigma = dp_i\wedge dx_i+
\displaystyle\frac{1}{2}\theta\,\epsilon_{ij}\,dp_i\wedge{}dp_j
+eB\,dx_1\wedge dx_2 + d\lf \frac{\vec{p}{\,}^2}{2m}+e V\rg \wedge
dt  , \label{Sourcoup}
\end{equation}
where $B\lf \vr\rg$ is the magnetic field, perpendicular to the
plane,  $V\lf \vr\rg$  the electric potential (both of them
assumed to be time-independent for simplicity) and $\theta$ is a
constant, called \textit{the non-commutative parameter}, for
reasons clarified below. The resulting equations of motion read
\cite{DH}
\begin{equation}
\displaystyle m^*\dot{x}_{i} = p_{i}-\displaystyle
em\theta\,\epsilon_{ij}E_{j},\qquad \displaystyle \dot{p}_{i} =
eE_{i}+eB\,\epsilon_{ij}\dot{x}_{j},
\label{DHeqmot}
\end{equation}
where  we have introduced the \textit{effective mass} $
m^*=m\;(1-e\;\theta \;B). \label{effmass} $ The novel physical
features are: i) the  \textit{anomalous velocity} term
$-\displaystyle em\theta\,\epsilon_{ij}E_{j}$, so that $\dot{x}_i$
and $p_{i}$ are not in general parallel, ii) the interplay between
the exotic structure and the magnetic field, yielding the
effective mass $m^*$ in (\ref{effmass}).

The \textit{LS} 2-form (\ref{Sourcoup}) can obtained by exterior
derivation from a  \textit{Cartan} 1-form on the evolution space
$\mathbb{{R}}^2 \times {\mathbb{{R}}}$, namely from
\begin{equation}
\lambda =  ({p_i}-{ A_i}\,) d{ x_i} -\frac{\vec{p}{\,}^2}{2m}\,dt
+ \frac{\theta}{2}\,\epsilon_{ij}\,{ p_i}\, d{ p_j}
\end{equation}
 defining the
action functional as in (\ref{1-order-action}). Accordingly,
 for $m^*\neq 0$, the associated Poisson
brackets are \begin{equation}
\begin{array}{lll}
\{x_{1},x_{2}\}= \displaystyle\frac{m}{m^*}\,\theta,   \quad
    \{x_{i},p_{j}\}=\displaystyle\frac{m}{m^*}\,\delta_{ij}, \quad
    \{p_{1},p_{2}\}=\displaystyle\frac{m}{m^*}\,eB.
\end{array}
\label{exocommrel}
\end{equation}
 and they automatically satisfy the Jacobi identity.
When effective mass vanishes, i.e. when the magnetic field takes
the critical value $ B_{crit}=\frac{1}{e\theta} $,  the system
becomes singular. Then symplectic reduction procedure  leads
to a two-dimensional system characterized by the remarkable
Poisson structure $ \{x_1,x_2\} = \theta $, reminiscent of the
``Chern-Simons mechanics'' \cite{DJT}. Thus, the symplectic plane
plays, simultaneously, the role of both configuration and phase
spaces. The only motions are those following the \textit{Hall law}.
Moreover, in the quantization of the reduced system, not only the
position operators no longer commute, but the quantized equation
of motions yields the \textit{Laughlin} wave functions \cite{QHE},
which are the ground states in the Fractional Quantum Hall Effect
(FQHE).   Thus one can claim that the classical counterpart of the
\textit{anyons} are in fact the\textit{ exotic} particles in the
system (\ref{DHeqmot}). In the review article \cite{HMS010}
several examples of 2-dimensional models, which generalize  the
form (\ref{Sourcoup}) and the equations (\ref{DHeqmot}) have been
discussed. Here let us conclude that the Poisson structure
(\ref{exocommrel})  can be obtained by applying the Lie-algebraic
Kirillov-Kostant-Souriau method for constructing dynamical
systems to the (2+1) 2-fold centrally extended Galilei group
\cite{Grigore,otherNC}. The two
 cohomological parameters are
the mass and a second,
 ``exotic'',  parameter, identified with a constant Berry curvature
 in the context of the condensed matter physics,
 or as a non relativistic limit of spin (see \cite{HMS010} and references therein).
 The latter is highlighted by
the non-commutativity of Galilean boost generators,
\begin{equation}
[K_1,K_2]=\imath \theta m^2 . \label{exorel}
\end{equation}
 \section{The  general model in (3+1)-dim}
After the above digression on two-dimensional models, let us look
for further generalizations \cite{BlochHam} - \cite{Blio1} of the
2-form (\ref{Sem2}) with \textit{momentum} dependent fields.
  Straightforward algebraic considerations  lead to   define on
  the space - momentum - time evolution space
 the manifestly anti-symmetric covariant  2-tensor  on the evolution space of
\beqa \sigma &=& \lq \lf 1 - \mu_{i}\rg d p_i - e\; E_i\; dt \rq
\wedge \lf dr_i - g_i\; dt \rg + \half \;e \; B_k \; \epsilon_{k i
j} \;dr_i \wedge dr_j  + \nn
\\ & & \half\;
\kappa_k\; \epsilon_{k i  j}\; d p_i \wedge dp_j +   q_k\;
\epsilon_{k  i  j}\; dr_i \wedge dp_j  , \label{Lag2Dsymm} \eeqa
 where we have put into evidence
    the Lorentz contribution to the Lagrangian,  the quantities
     ${\vec g}$, $\vec \kappa$ and  $\vec q$ are
 3-vectors  and $Q = \textrm{diag}\lf \mu_{i} \rg $ is a $3\times 3$ matrix, respectively. They depend
on all independent variables and have to be determined in such
a way that $\sigma$ is closed and has constant rank.  By the
 expression  $ 1 - \mu_{i}$  we would like  to distinguish among
the  \textit{bare}  mass, normalized to 1, and the possible
variable contributions.
 It is interesting to note that, with
respect to the expressions adopted in \cite{SSD} and \cite{DH}, we
have to introduce the 1-form $dr_i - g_i\; dt $, which defines a
new conjugate momentum.

The equations of motion can be written as the kernel of the
\textit{LS} 2-form $\sigma \lf \delta y, \cdot \rg =0 $ for a
tangent vector $\delta y = \lf \delta \vr, \delta \vp, \delta t
\rg $. Specifically one obtains the equations \beqa
    e \;\delta \vr  \cdot \vE -  \lf 1- Q \rg  \delta \vp \cdot {\vg} &
= &0,\label{eqEn} \\
  \lf 1- Q -Q_A\rg \delta \vp
& = & e \lf  \vE \; \delta t + \;  \delta \vr \times \vB \rg, \label{eqLor} \\
  \lf 1- Q \rg \;\lf \delta \vr - {\vec g}\; \delta t \rg & = & {
 -\delta \vr \times \vec q} \;-    \delta \vp \times {\vec \kappa} \;,
\label{eqParEx} \eeqa where we have introduced the anti-symmetric
matrix $\lf Q_A \rg_{i j} = \epsilon_{i j k}\; q_k$. From the
equation (\ref{eqLor}) we can solve with the respect of the vector
$\delta \vp$, if the mass matrix $M = {\mathbf{1}} - Q - Q_A$ is
invertible, i.e. if ${\rm det}\lf M \rg  \neq 0$. Under such an
hypothesis together with ${\rm det}\lf m-Q \rg  \neq 0$, one
   replaces $\delta \vp$ in the equation (\ref{eqParEx}), finding
   an equation for the position tangent vector $\delta \vr$ of the form
   \beq  M^* \; \delta \vr = \lf \lf 1- Q\rg {\vec g} + e\; {\cal K} M^{-1} \cdot \vE   \rg \delta
   t, \label{RMotion}
   \eeq where the  effective mass  matrix  is given by \beqa
   M^{\star}=  M + \lf 2 Q_A - e\; { \Theta} M^{-1} {\cal B} \rg  \; ,
   \quad { \Theta}_{i j} = \epsilon_{i j k} \kappa_k\;, \;
    {\cal B}_{i j} = \epsilon_{i j k}
   B_k. \label{EffMass}
    \eeqa
    Both $M^*$ and the eq.  (\ref{RMotion})
   generalize of the expressions obtained in \cite{DH} leading to (\ref{DHeqmot}).
   Singularities in the motion can arise, both from the vanishing of ${\rm det}\lf M \rg$
and degeneracies of the first factor in (\ref{EffMass}), that is
at ${\rm det}\lf M^*\rg = 0$.
   However, if this is not the case,
one can solve (\ref{RMotion}) w.r.t $\delta
   \vr$ and show that
\begin{enumerate}
\item equation (\ref{eqEn}) is identically  satisfied independently
from the specific choice for the vector $\vec g$,
\item the equation (\ref{eqLor}) becomes
\beq
 M^*\delta\vp =
\frac{e}{{\rm Det}\lf M\rg}\lf  R\; \vE - {\vec g}^{\; T} N {\vB}
\rg \delta t, \label{eqLorMod} \eeq where matrices $ R $ and $N$
have an involved dependency on $m$, $\vB$, $\vec \kappa$, $Q$ and
$Q_A$ to be spelled here. \end{enumerate}
 Thus we have a system of simultaneous first order differential equations:  (\ref{RMotion})
for the velocity of the particle $\frac{\delta \vr}{\delta t}$ and
(\ref{eqLorMod}) for
   the momentum variation $\frac{\delta \vp}{\delta t} $. Notice that these two equations simplify to
the equations (8) and (9) of the \cite{DH}, when $Q$ and  $Q_A$ go
to 0 and $ {\vec g} \equiv  \vp$.

Now,  the closure condition in the evolution space $ d\sigma = 0 $
( called also ``Maxwell Principle'' by Souriau \cite{SSD}) of the
Lagrangian-Souriau form $\sigma$ in (\ref{Lag2Dsymm}) leads to the
coefficients of the independent $3$-forms. It is quite natural to
assume the following limitations on the involved functions:
$\p_{p_i} E_j = \p_{p_i} B_j =0 $ . Thus we are  lead to the
equations
 \beqa \p_{r_j} B_j  = 0,  &  \varepsilon_{k i j } \p_{r_i} E_j = -
\p_t  B_k \label{MaxwellHom}
 , \\ \p_{p_j}  \kappa_j  = 0 ,   & \varepsilon_{k i
j} \p_{p_i} \lq \lf 1- \mu_j \rg g_j \rq =  \p_t \kappa_k,\label{eqtk}\\
 \p_t \mu_{i} =  \p_{r_i} \lq \lf 1- \mu_i \rg g_i \rq  ,   &
 \half \varepsilon_{k i j}\;\p_{r_i} \, \lq \lf 1- \mu_j \rg g_j \rq =  \p_t q_k  ,\label{eqtm}
\\ \p_{r_i} \;\mu_{j} =  \varepsilon_{i j k} \p_{r_j} q_k , &
\p_{r_i} \kappa_j = \varepsilon_{i j k}\p_{p_k} \; \mu_{i} +
\p_{p_i}\; q_j - \delta_{i j} \p_{p_k}\; q_k , \label{skewdiagr2}&
\eeqa
                           with  the residual closure relations
                           (without summation  over repeated indices)
 \beq  \p_{r_j} \lq \lf 1- \mu_i \rg g_i \rq + \p_{r_i} \lq \lf 1- \mu_j \rg g_j \rq = 0, \qquad
i\neq j =1, 2, 3.\label{cond_momento}\eeq One can observe that the
homogeneous Maxwell equations (\ref{MaxwellHom}) are the only
restrictions on the electromagnetic fields $\lf \vE, \vB \rg$.
Equations (\ref{eqtk}) are the analogs of the previous relations
in the velocity (momentum) space for the vector-field $\vec
\kappa$, which measures the extent to which the spatial
coordinates fail to commute in three dimensions, as we will see in
the Hamiltonian formalism. For such a reason, sometimes $\vec
\kappa$ is called dual magnetic field.

 If $\vec \kappa$ is non
trivial in time, then a change in the velocity dependence is
induced  for the mass flow $  \lf \mathbf{1}- Q \rg \vg $, as
prescribed by the second equation in (\ref{eqtk}). In its turn,
equations (\ref{eqtm}) say how the particle mass may change in
time. This seems to be a quite unusual situation, but we cannot
discard it at the moment.  On the other hand,  the first set of
three equations in (\ref{eqtm}) has the form of independent
continuity equations, leading to the global conservation law for
the total mass, i.e. $\p_t \lf \sum_i \mu_i\rg + \p_{r_i} \lq \lf
1- \mu_i \rg g_i \rq = 0 $, which however holds separately in
different directions. Due to $\vq$, also the skew-symmetric
contributions to the mass matrix $M$ may change on time, but  they
generate modification of the mass flux in space, accordingly to
the  second set   of equations in (\ref{eqtm}).

 The equations in
(\ref{skewdiagr2}) are more difficult  to interpret: they provide
consistency relations for both the space and the velocity
dependency among the mass matrix elements and the dual magnetic
field. Putting such expressions into the equation of motion in the
form (\ref{eqLor})-(\ref{eqParEx}),  in a pure axiomatic way one
re-obtains the equations found in the context of the semiclassical
motion of electronic wave-packets in \cite{Niu}.

 For a  particle in a \textit{flat} evolution space, i.e. for $\mu_i \equiv 0$ and momentum
${p_i}= g_i $, one easily concludes that $\vec \kappa$ and $\vq$
have to be  constants.    The analysis of the closure relations
(\ref{eqtm})-(\ref{skewdiagr2}) leads to the expression $\kappa_i
= \sum_{j\neq i} \lf x_j \p_{p_j} q_i -x_i \p_{p_j} q_j \rg +
\chi_i$ ,where the $q_i$'s and   $\chi_i$ 's depend only on the
velocities and moreover the divergenceless condition $\p_{p_j}
\chi_j = 0 $    has to be satisfied.  A remarkable example for its
phenomenological implications is provided by the monopole field in
momentum space ${\vec \kappa} = \theta \frac{\vp}{|\vp|^3}$, which
is indeed the only possibility consistent with the spherical
symmetry and the canonical relations $\{r_i,p_j\}=\delta_{ij}$
\cite{BeMo}. Its expression  appears to be consistent, at least
qualitatively, with the data reported in (\cite{AHE}) and in Spin
Hall Effects \cite{SpinHall}, \cite{HorvMonop}.

Limiting ourselves to two spacial dimensions and setting
$\kappa_3=\const$, a free particle admits in fact the ``exotic''
two-parameter centrally extended Galilei group as symmetry
\cite{DH}. In a more general situation one deals with a
momentum-dependent non-commutativity  field
$\kappa_3=\kappa(\vp)$,  like it was considered by \cite{Snyder}
with  $\kappa = -\frac{\theta}{1 + \theta p^2} \varepsilon_{i j}
p_i r_j$  (and $\mu_i \equiv 0$ ), or  $ \kappa^{\alpha \beta} =
\frac{s}{2}
 \frac{p_\alpha \epsilon^{\alpha\beta\gamma}}
 {(p^2)^{3/2}}$  for the planar relativistic model for a  spinning
 particle (again a sort of monopole in relativistic momentum
 space) \cite{anyoneqs,HP1}.

  Finally, in the singular submanifold of the
   phase space defined by $
 M^*=0
$,
 we need to look at the proper restrictions on
   vector-fields ${\delta \vp}$ and ${\delta \vr}$, in order to
   avoid motion with infinite velocities. In particular in two dimensions, assuming that the only
   non vanishing components of the magnetic, of the dual magnetic fields
   and of the $\vq$
    are the ones   orthogonal to the plane of motion, with ${\rm det}\lf 1- Q \rg \neq
    0$,
   the singularity mass manifold is described by
\beq
   B_{crit} = \frac{q^2+\left(m-\mu_1\right)
   \left(m-\mu_2\right)}{e \kappa }.
 \eeq    For sake of simplicity, here we assume all the above quantities as constants
 on space-velocity variables.   The equations of motion
 (\ref{eqEn}) - (\ref{eqParEx}) and the closure relations are
 satisfied by the following constraints  for the mass flow and
 electric field components
 \beqa  \lf 1- \mu_i \rg g_i = -\frac{e \kappa  \left( q
   E_i- \epsilon
   _{i j} E_j
   \left(  1- \mu_i \right) \right)}{q^2+\left( 1- \mu_1\right)
   \left( 1- \mu_2\right)} = \frac{\left( q
   E_i- \epsilon_{i j} E_j
   \left( 1- \mu_i\right) \right)}{B_{crit}} ,  \label{genHall}   \\
  \left( 1- \mu_2 \right)
    \p_{r_1} E_1  + \lf \mu_1 - 1\rg
   \p_{r_2} E_2 + q
   \left( \p_{r_1} E_2    + \p_{r_2}  E_1\right) = 0.
\label{ElectricConstraint} \eeqa    Notice
   that the last relation comes from the closure condition on the
   mass terms (\ref{cond_momento}), not from the Faraday law
   (\ref{MaxwellHom}), which is identically satisfied. Since the same equations provides the time
   evolution of the mass terms and of $\kappa$, the only left by
   above assumptions, the electric field can depends at most
   linearly on space coordinates, compatibly   with
   (\ref{ElectricConstraint}).
   Finally, the equation (\ref{genHall})  generalizes the
    {\it Hall law} discussed in the previous Section.


\section{Hamiltonian Structure}

Comparing the system (\ref{RMotion})-(\ref{eqLorMod}) with the
previous ones in (\ref{DHeqmot}), one recognizes the general
common structure, due to derivation from the same unifying
Lagrange-Souriau approach provided by the 2-form
(\ref{Lag2Dsymm}), even if the peculiarity of the doubled folded
central extensions is enjoyed only in the 2-dimensional setting.

The 2-form $\sigma $ can be obtained as the exterior
derivative of the Cartan 1-form,
\beq \lambda = \lf \vp +
 \overrightarrow{{\cal A}} \rg \cdot d \vr + {\vec{\cal R}}\, \cdot d \vp -  {\cal T}\, dt.
 \label{cartan1}
\eeq  In this formula the field     $\overrightarrow{{\cal A}} \lf
\vr, t \rg$ is the usual electromagnetic potential, such that $\vB
= \nabla_{\vr} \times \overrightarrow{{\cal A}}$, for which we
have postulated only a space - time dependency, eventually
resorting to a suitable gauge transformation. Then, the electric
field (in fact any space-time dependent force), is given by $\vE =
- \nabla_{\vr} {\cal T} - \p_t \overrightarrow{{\cal A}} $, where
we assume the following decomposition for the scalar function
${\cal T}\lf \vr, \vp, t \rg = {\cal E}\lf \vp, t \rg+ \varphi\lf
\vr,  t \rg$, in order to keep valid the previous restrictions on
the dependency of  $\vE$ fields. On the other hand, the field
${\vec{\cal R}} \lf \vr, \vp, t \rg $ defines the dual magnetic
field $\vec{\kappa} = \nabla_{\vp} \times {\vec{\cal R}} $, the
mass flow is given by $\lf \mathbf{1} - Q \rg {\vec g} =
-\nabla_{\vp} {\cal T} - \p_t {\vec{\cal R}} $, $\mu_i = \p_{r_i}
{\cal R}_i $ and $q_k = \p_{r_i}{\cal R}_j = -
 \p_{r_j}{\cal R}_i$, where $k, i, j$ are in cyclic order.
 The last equality imposes a set of constraints on
${\vec{\cal R}}$ in such a way the Lagrangian-Souriau 2-form takes
exactly the expression (\ref{Lag2Dsymm}). The above restrictions
implies certain second order relations, which assure that also the
first set of closure relations in (\ref{skewdiagr2})     are
satisfied, namely  \beq \p_{r_i}\p_{r_j} {\cal R}_k =0 \quad
\textrm{( $i, j, k$ cyclic)},\qquad \p_{r_i}^2  {\cal R}_j = -
\p_{r_i}\p_{r_j}  {\cal R}_i \;\; \lf i \neq j \rg
,\label{constraints}\eeq with no summation over repeated indices
in the last equation. The remaining closure relations in
(\ref{skewdiagr2}) are identically satisfied, while those in
(\ref{eqtm}) require the supplementary constraints on time
derivatives \beq \p_t \lf \p_{p_i}{\cal R}_j + 2 \p_{p_j}{\cal
R}_i  \rg = 0  \qquad i \neq j.\eeq Due to the special form we
assumed on the force and magnetic fields, the above restrictions
on $\vec{\cal R}$ limit its space-time dependency, leaving however
the gauge freedom with respect the momentum variables.

Thus, in terms of the potentials, or connections, introduced in
(\ref{cartan1}), the  equations of motion
(\ref{eqLor})-(\ref{eqParEx}) become
 \beqa \lf \delta_{i j} + \Xi_{i j} \rg\, \dot{r}_j  +
\Theta_{i j}\,  {\dot{p  }}_j  & = &
 \p_{v_i  }  \cE
 + \p_t {\cR}_i, \nn
\\
 {\cal B}_{i j}\, {\dot{r}_j
} + \lf \delta_{i j} + \Xi_{i j}  \rg\,\dot{p}_j   & = & -\p_{r_i}
\varphi  -\p_t { \cA}_i, \label{eqmotgen} \eeqa
 where we have used the
matrices (see also eq. (\ref{EffMass})) \beqa \Xi_{i j} =\left\{%
\begin{array}{ll}
     - \p_{r_j} \cR_{i}, & i \leq j \\
  \p_{r_i} \cR_{j} , & i > j \\
\end{array}%
\right.    \quad {\cal B}_{i j} = \varepsilon_{i j k }
\varepsilon_{j h k } \p_{r_h}\cA_k , \quad \Theta_{i j} =
\varepsilon_{i j k } \varepsilon_{j h k } \p_{p_h}\cR_k .
 \label{altriTens} \eeqa
Differently from system (\ref{eqLor})-(\ref{eqParEx}) on the
evolution space, now the dynamical system (\ref{eqmotgen}) is
defined on the tangent manifold of the phase space $TQ \equiv \lgr
\xi = \lf \vr , \vp \rg \rgr$ space.  But, if $\p_t \vec{\cA} =
\p_t \vec{\cR} \equiv 0 $, it is possible to rearrange the
 form (\ref{Lag2Dsymm}) as  $
\sigma=\omega-dH\wedge dt $  by introducing the symplectic 2-form
on $ TQ$ \beq \omega = \lf \delta_{i,j}+ \Xi_{ij}\,\rg d r_{\;i}
\wedge d p_{\;j} + \frac{1}{2} \lq {\cal B}_{i j} \, d r_{\;i}
\wedge d r_{\;j} - \Theta_{i j} \, d p_{\;i} \wedge d p_{\;j} \rq
\label{TQHam} \eeq
 and the Hamiltonian function $ { H}\equiv {\cal T} = {\cal E}\lf \vp, t \rg+
\varphi\lf \vr,  t \rg  $.

Thus, the closure of $\omega = \omega_{\alpha \beta}\, d
\xi_{\alpha} \wedge d \xi_{\beta}$ is assured by that one of
$\sigma$,  i.e. by the equations (\ref{MaxwellHom}) -
(\ref{cond_momento}) plus the restriction (\ref{constraints}).
Then, the space $TQ$ becomes a Poisson manifold, with  Poisson
brackets defined,  as in (\ref{Poisson}),  by the co-symplectic
matrix  \beqa & \omega^{\alpha, \beta} = \Big(1 - \frac{1}{2} {\rm
Tr} \lf \Xi ^2 + X \lf {\bf 1} + 2 \, \Xi \rg \Theta \rg\Big)^{-1}
\\
 &  \lgr \lf {\begin{array}{cc} \Theta + \lq  \Xi,
\Theta\rq & 0
 \\
0  & - {\cal B} + \lq \Xi, {\cal B} \rq \\
\end{array}}
\right) +      \lq 1 - \frac{1}{6}
 {\rm Tr } \lf \Xi^2 + {\cal B} \, \Theta \rg \rq
 \lf {\begin{array}{cc} 0 &  1
 \\
- 1 & 0 \\
\end{array}}     \nn
\right) + \right. \\ &\left. \lf {\begin{array}{cc} 0 &  \lf \Xi^2
+ X \,
 \Theta \rg^T
 \\
 -  \lf  \Xi^2 + {\cal B} \, \Theta \rg  & 0 \\
\end{array}}
\right)\rgr,   \nn
 \label{cosympl} \eeqa
  non degenerate for
$ \sqrt{\det\lf \omega_{\alpha\beta}\rg} =  1-\frac{1}{2}{\rm Tr}
 \lf \Xi^2 + {\cal B} \lf {\bf 1} +
2 \, \Xi \rg \Theta \rg  \neq 0 . $ Such a factor generalizes the
denominators present in the Poisson brackets (\ref{exocommrel}) or
(\ref{EffMass}). Moreover, it crucially appears in the expression
of the invariant phase - space volume, ensuring the validity of
the Liouville theorem.  Finally, notice that the Poisson structure
is determined only by gauge invariant quantities and  brackets
involving position coordinates $r_i$ are in general
non-commutative. As it has been elsewhere remarked \cite{HMS010},
it is possible to perform a change of variables leading to
commutative position variables by a point transformation of the
form $r_i \to r'_i = r_i - {\cal R}_i ({\vr}, {\vp})$. However,
the vector field $\vec {\cal R}$ is defined up to a gauge
transformation generated by an arbitrary function on $\lf \vr, \vp
\rg$. Thus, the  meaning of the notion of position  is unclear in
such a context.

For a charge subjected only  to a monopole of strength $\theta$ in
momentum space and to a uniform electric field, the equations of
motion obtained (\ref{cosympl}) are readily integrated
\cite{HorvMonop}. The particle suffers a shift $\Delta  = \frac{2
\theta}{ p_0} $ in the direction $\vE \times {\vec p}_0$, being
${\vec p}_0$ the initial  linear kinetic momentum. This quantify
the discussion in point b) in the Introduction. On the other hand,
if the charge of Hamiltonian $H = \frac{|\vp|^2}{2}$ is driven
only by a magnetic and by a dual monopole, the equations are \beqa
\frac{M^*}{|\vr|^3 |\vp|^3} \dot{r}_i &=& p_i - e \theta
\frac{r_i}{|\vp| |\vr|^3}, \;
\\
 \frac{M^*}{|\vr|^3 |\vp|^3} \dot{p}_i &=& e \varepsilon_{i j k} \frac{p_j r_k}{|\vr|^3}.
\eeqa where $ M^* =  |\vr|^3 |\vp|^3 - e \theta \;\vr \cdot \vp $.

A final remark in connection with the coupling to the
electromagnetic field adopted in (\ref{Lag2Dsymm}), sometime said
\textit{the minimal addition} and leading to the Hamiltonian
${\cal T}$ above. It is quite different from the usual
\textit{minimal coupling}  procedure and it  yields a very
different Poisson structure. In the context of the 2-dimensional
systems this problem was reviewed in \cite{HMS010}. In particular
the two formulations, in that context, were proved to be
equivalent under a classical Seiberg-Witten transformation of
electromagnetic fields. No results are yet available in the
situations discussed in the present paper.

In conclusion a wide set of dynamical systems is derived from the
Lagrange-Souriau approach in 3-dimensions. Generalizations to
higher number of degrees of freedom seems straightforward.  We
have shown the conditions to assure their Hamiltonian formulation.
From which an analysis for their integrability properties can be
pursued more plainly, by resorting to standard methods.
Alternatively, one can perform a symmetry analysis directly on the
\textit{LS} 2 -form  (\ref{Lag2Dsymm}) accordingly to
\cite{Grigore}and \cite{Banerjee}.

\section*{Acknowledgments}
 The author expresses his indebtedness to P. Horvathy for having been introduced to the problem and
 for continuous and stimulating discussions.
Some of the   results were obtained in collaboration with C. Duval, Z. Horv\'ath and
 P. Stichel. The work has been partially supported
 by the INFN - Sezione of Lecce under the project LE41.


 \end{document}